\documentclass[english,aps,twocolumn,prd,superscriptaddress]{revtex4-1}

\usepackage{enumerate}

\usepackage{amsmath,amstext,amsfonts,amscd,latexsym,amssymb,amstext,mathrsfs,dsfont,xfrac}

\usepackage{color,amsmath,amssymb,exscale,epsfig}
\usepackage{hyperref}

\usepackage[none]{hyphenat}

\usepackage{graphicx,color,pst-plot,axodraw4j}
\usepackage{pgfplots,tikz}

\allowdisplaybreaks[1]

\def\R{{\rm Re\,}}
\def\I{{\rm Im\,}}

\def\d{{\rm d }}
\newcommand{\hyper}[5]{\;_{#1}{\rm F}_{#2} \left(\left.\begin{matrix} {#3} \\ {#4} \end{matrix}\right| #5\right) }
\newcommand{\braket}[3]{\left<\; #1 \;\left|\; #2 \;\right| #3\;\right>}

\begin{document}

\title{$K \rightarrow \pi \ell^+ \ell^-$ form factor in the Large-N$_c$ and cut-off regularization method}

\author{Estefania Coluccio Leskow}
\email{coluccio@na.infn.it}
\affiliation{INFN-Sezione di Napoli, Via Cintia, 80126 Napoli, Italia}

\author{Giancarlo D'Ambrosio}
\email{gdambros@na.infn.it}
\affiliation{INFN-Sezione di Napoli, Via Cintia, 80126 Napoli, Italia}

\author{David Greynat}
\email{greynat@na.infn.it}
\affiliation{INFN-Sezione di Napoli, Via Cintia, 80126 Napoli, Italia}
\affiliation{Dip. di Scienze Fisiche, Universit\'a di Napoli "Federico II", Via Cintia, 80126 Napoli, Italia}

\author{Atanu Nath}
\email{nath@na.infn.it}
\affiliation{Dip. di Scienze Fisiche, Universit\'a di Napoli "Federico II", Via Cintia, 80126 Napoli, Italia}

\begin{abstract}
Bardeen-Buras-G\'{e}rard \cite{Buras:1985yx,Bardeen:1986vp,Bardeen:1986uz,Bardeen:1986vz,Hambye:1998sma,Gerard:1990dx,Buras:2014maa} have proposed a large N$_c$ method to evaluate hadronic weak matrix elements to attack  for instance the determination of the $\Delta I= \sfrac{1}{2}$-rule and $\R  (\sfrac{\epsilon'}{\epsilon})$. Here we test this method to the determination of the form factor parameters $a_+$ and $b_+$ in the decays $K^+ \rightarrow \pi^+ \ell^+ \ell^- $ and $K_S \rightarrow \pi ^0 \ell^+ \ell^-$. The results are encouraging: in particular after a complete treatment of Vector Meson Dominance (VMD).

\end{abstract}

\maketitle

\section{Introduction}
Rare kaon decays play a crucial role in particle physics~\cite{deRafael:1995zv,Pich:1995bw,Ecker:1995zu,D'Ambrosio:1996nm,Komatsubara:2012pn,Cirigliano:2011ny,Littenberg:2002um,Colangelo:2000zw} particularly now with the  beautiful physics program of NA62~\cite{NA62}, where 100 events of $K^+ \rightarrow \pi^+ \nu \bar{\nu}$ are expected and  the J-PARC KOTO experiment with the goal of  a few  $K_L \rightarrow \pi ^0 \nu \bar{\nu}$  SM events in 3-4 years run with Signal/Noise ratio $\sim 2$~\cite{Ahn:2009gb}. 
 
Similar to $K_L \rightarrow \pi ^0 \nu \bar{\nu}$, the short-distance (SD) part of $K_L \rightarrow \pi^0 e^+ e^- $ gives information on $V_{ts}V_{td}^*$ and thus measures the height of the unitarity triangle. The measurement of this decay may also lead to New Physics test \cite{Smith}. There is also an indirect CP-Violating contribution from $K_S \rightarrow \pi ^0 e^+ e^- $,  the magnitude of which can be obtained from the measured BR for the corresponding $K_S$ decay \cite{NA48/1,Andreazza:2015bja}. Also a theoretical determination is needed  and the recent lattice RBC and UKQCD Collaborations \cite{lattice} address this issue. The related $K^\pm \rightarrow \pi^\pm \ell^+\ell^-$ decay  may help also to this  goal; the experimental form factor here has been measured  well  by NA48/2 \cite{NA48/2,Andreazza:2015bja,NA62}. The appearance of chiral unknown constants \cite{Kpee_LD,D'Ambrosio:1998yj} brings up the crucial question to determine them either by lattice \cite{lattice} or in a model dependent manner \cite{deRafael:1995zv,Pich:1995bw,Ecker:1995zu,D'Ambrosio:1996nm,Komatsubara:2012pn,Cirigliano:2011ny,Littenberg:2002um,Colangelo:2000zw,D'Ambrosio:2002fa} as we will do in this paper. Since one can measure  $K^\pm \rightarrow \pi^\pm e^+ e^- $ and $K^\pm \rightarrow \pi^\pm \mu^+ \mu^-$ separately, the question of Lepton Flavor Universality Violation  is also interesting \cite{Sierra:2014nqa,Heeck:2014qea,Crivellin:2015mga,Dorsner:2015mja,Omura:2015nja,Varzielas:2015joa,Buras:2013dea, Crivellin:2015lwa,Glashow:2014iga,Altmannshofer:2014cfa,Gauld:2013qja,Buras:2013qja,Gauld:2013qba,Descotes-Genon:2013wba,Niehoff:2015bfa,Sierra:2015fma,Crivellin:2015era,Celis:2015ara,Carmona:2015ena,Varzielas:2015iva,Becirevic:2015asa,Gripaios:2014tna,Calibbi:2015kma,Alonso:2015sja,
Bauer:2015knc,Barbieri:2015yvd,Crivellin:2012ye,Tanaka:2012nw,Celis:2012dk,Crivellin:2013wna,Crivellin:2015hha,Fajfer:2012jt,Deshpande:2012rr,Sakaki:2013bfa,
Fajfer:2015ycq,Greljo:2015mma,Buras:2015xba,Buras:2015yba,Boucenna:2015raa,Crivellin:2016vjc}.

In this paper we will evaluate the $K^\pm \rightarrow \pi^\pm \ell^+ \ell^-$ form factor in the theoretical framework suggested by Bardeen-Buras-Gerard (BBG) \cite{Buras:1985yx,Bardeen:1986vp,Bardeen:1986uz,Bardeen:1986vz,Hambye:1998sma,Gerard:1990dx,Buras:2014maa}; the authors of this approach have successfully applied the method to the explanation of the $\Delta I=\sfrac{1}{2}-$rule and $\pi^+-\pi^0$-mass difference: we think it is interesting to apply it here. 
 
The recent lattice  result from RBC and UKQCD Collaborations~\cite{lattice} reporting on the $K\to \pi \pi$ matrix element $\R(A_0)$ and $\I(A_0)$ leading to $2$--$3\sigma$ below the experimental world average of $\R(\sfrac{\epsilon^\prime}{\epsilon})$ has led the authors of Refs.~\cite{Buras:2015xba,Buras:2015yba} to evaluate the same weak matrix elements $B_6$ and $B_8$ in their approach, finding consistency with lattice results and they conclude~\cite{Buras:2015yba} that New Physics seems required to accommodate  the present  experimental value of $\R(\sfrac{\epsilon '}{\epsilon})$. Using Large N$_c$ and Minimal Hadronic Ansatz,  Hambye \textit{et al.} in Ref.~\cite{Hambye:2003cy} still find agreement with experimental values.  Final state interaction is not accurately described by lattice (the lattice result ~\cite{lattice} for the $I=0$ phase-shift $\delta_0=23.8(4.9)(1.2)^\circ$ is about $3\sigma$ smaller than the value obtained in dispersive treatments of Ref.~\cite{Colangelo:2001df,GarciaMartin:2011cn,Colangelo:NA62}) and a good theoretical description  could lead to agreement with experiments as the approach of Refs. \cite{Pallante:1999qf,Pallante:2000hk}; for an alternative solution see Ref. \cite{Buras:2016fys}.

Nevertheless we think it is interesting to check BBG method in $K^+ \rightarrow \pi^+ \ell^+\ell^-$ decay.  We dedicate section II and III to model independent discussion, section IV to the BBG method, V to the form factor evaluation, VI to the addition of vectors and VII to the $K_S\rightarrow \pi ^0 \ell^+\ell^-$-form factor.

\section{Model independent analysis}

The decay $K\rightarrow \pi \bar{\ell}\ell$ is dominated by a virtual photon exchange \cite{Kpee_LD,D'Ambrosio:1998yj},
\begin{multline}
\mathcal{A}\left[K(k) \rightarrow \pi(p) \gamma^*(q) \right] \\
= \frac{W_+(z)}{(4\pi)^2}\left[z(k+p)_\mu - (1-r_\pi^2)q_\mu\right]\label{eq:defAmp},
\end{multline}
where $r_x \doteq \frac{M_x}{M_K}$ and $z \doteq \frac{q^2}{M_K^2}$, with $q^2$ being the photon transferred momentum.  With these conventions the decay amplitude takes the form ($\alpha\doteq \sfrac{e^2}{4\pi}$),
\begin{multline}
\mathcal{A}\left[K(k) \rightarrow \pi(p) \ell^+(p_+) \ell^-(p_-) \right] \\= -\frac{\alpha}{4\pi M_K^2}\;W_+(z)\;(k+p)_\mu\; \bar{u}_\ell(p_-) \gamma^\mu  v_\ell(p_+)\;.\label{eq:amplitude}
\end{multline}

The form factor $W_+(z)$ can be decomposed into two parts: one coming from the dominant pion loop contribution  $W_+^{\pi\pi}(z)$, and another one $W_+^{\rm pol}(z)$,  that accounts for the contributions of higher mass intermediate states (like $K^+ K^-$ for instance and local pieces).  $W_+^{\rm pol}(z)$ can be well approximated by a linear polynomial for small values of $z$,  $W_+^{\rm pol}(z) \sim a_++b_+z$. In this way, $W_+(z)$ can be written as~\cite{D'Ambrosio:1998yj},
\begin{equation}
\label{eq:W+decomp}
W_+(z) = G_F M_K^2 (a_++b_+z) + W_+^{\pi\pi}(z)\;,
\end{equation}  
with a priori unknown low-energy constants contributing to $a_+$ and $b_+$ which have to be experimentally determined \cite{NA48/1,NA48/2,NA62} .
 
\begin{figure}[h]
\begin{center}
\scalebox{0.5}{   \begin{picture}(228,151) (63,-43)
    \SetWidth{1.0}
    \SetColor{Black}
    \Line[arrow,arrowpos=0.5,arrowlength=5,arrowwidth=2,arrowinset=0.2](80,23)(128,23)
    \Arc(160,23)(32,180,540)
    \Line[arrow,arrowpos=0.5,arrowlength=5,arrowwidth=2,arrowinset=0.2](128,23)(128,87)
    \Photon(192,23)(256,23){7.5}{3}
    \Text(60,23)[lb]{\Large{\Black{$K$}}}
    \Text(112,87)[lb]{\huge{\Black{$\pi$}}}
    \Text(256,23)[lb]{\huge{\Black{$\gamma^*$}}}
    \SetColor{Gray}
    \GOval(128,23)(16,16)(0){0.823}
    \Text(176,55)[lb]{\huge{\Black{$\pi$}}}
    \Text(176,-25)[lb]{\huge{\Black{$\pi$}}}
    \SetColor{Black}
    \Line(160,71)(161,-42)
    \Line(161,-42)(169,-33)
    \Line(160,-25)(168,-16)
    \Line(160,-9)(168,0)
    \Line(160,7)(168,16)
    \Line(160,23)(168,32)
    \Line(160,39)(168,48)
    \Line(160,55)(168,64)
    \Line(160,71)(168,80)
  \end{picture}
}
\end{center}
\caption{Pion loop contribution to $K^+\rightarrow \pi^+\gamma^{*}$. The unitary cut used is represented too. The blob represents the $K \rightarrow 3\pi$ vertex.}
\label{fig:1}
\end{figure}
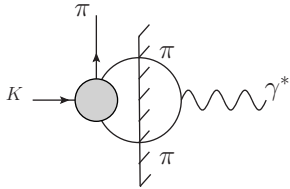

$W_+^{\pi\pi}(z)$ is obtained from the analytic structure of the diagram in Fig.~\ref{fig:1}~\cite{D'Ambrosio:1998yj}. 

In Ref.~\cite{D'Ambrosio:1998yj}, the behaviour of $W_{+}(z)$ at $z \rightarrow 0$ is entirely fixed up to $W_+^{\rm pol}(z)$,
\begin{multline}
\label{eq:Wexpansion}
W_+(z) \underset{z \rightarrow 0}{\thicksim}  G_F M_K^2 a_+ \\
+ \left( G_F M_K^2 b_+ + \frac{3r_\pi^2(\alpha_+ - \beta_+) - \beta_+}{180 r_\pi^6}  \right) z \;,
\end{multline}
where $\alpha_+ = (-20.6 \pm 0.5) \cdot 10^{-8}$ and $\beta_+ = (-2.6 \pm 1.2) \cdot 10^{-8}$ are the $K\rightarrow 3\pi$ parameters from Ref. \cite{Devlin:1978ye,Kambor:1991ah,Kambor:1992he,Bijnens:2004ku,Bijnens:2004vz,Bijnens:2004ai}. 

The local counter-term structures at $\mathcal{O}(p^4)$ are
\begin{equation}
a_+^{(4)} = \frac{G_8}{G_F}\left( \frac{1}{3}-w_+\right)\;,
\end{equation} 
where $w_+$ is given \cite{Kpee_LD} in terms of $N_i$'s \cite{Ecker:1992de} and $L_9$ \cite{Gasser:1984pr}  by 
\begin{equation}
\label{eq:w+}
w_+  =  \frac{64 \pi^2}{3} \left( N_{14}^r - N_{15}^r + 3 L_9^r \right) + \frac{1}{3} \ln \frac{\mu^2}{M_K M_{\pi}}\;.
\end{equation}
Since $w_+$ is scale independent, the $\mu$ dependence of the combination among the $N_i$'s and $L_9$ is exactly compensated by the $\log \mu^2$ from the chiral loop
in eq. \eqref{eq:w+}. From the loop contribution, one has also, 
\begin{equation}
b_+^{(4)} =- \frac{G_8}{G_F} \frac{1}{60}\;.
\end{equation} 

Experimentally \cite{NA48/2}, we have 
\begin{align}
a_+^{\text{exp.}} &=- 0.578 \pm  0.016\;, \\
b_+^{\text{exp.}} &=- 0.779 \pm  0.066 \label{eq:exp}\;. 
\end{align}

As we can see, the experimental values for $a_+$ and $b_+$ are the same order of magnitude. We then have to understand why so. Indeed, $a_+$, $L_9$ and the $N_i$'s have large contributions from the VMD \cite{Ecker:1992de,D'Ambrosio:1997tb,Cappiello:2011re} , since $b_+$ is mostly a $\mathcal{O}(p^6)$ observable, it should have an important enhancement. 

The $K_S$ decay is discussed in Sec. \ref{sec:KS}.

\section{Amplitude analysis and short distance results}

The behavior of the amplitude in eq.~\eqref{eq:amplitude} can be studied distinguishing two different contributions: \textit{(i)} the long-distance (LD) one described by chiral perturbation theory ($\chi$PT) \cite{Kpee_LD,D'Ambrosio:1998yj}, and \textit{(ii)} the short-distance (SD) one described by an effective four-quark Hamiltonian \cite{Gilman_Wise,Buras:1991jm,Buras:1992tc,Buras:1992zv,Buchalla:1995vs,Buras:1998raa}. The complete description of the amplitude implies then a continuation through both regions.
  
The dominant $\Delta S=1$, SD effective four-quark Hamiltonian is given by \cite{Gilman_Wise,Buras:1991jm,Buras:1992tc,Buras:1992zv,Buchalla:1995vs,Buras:1993dy,Buras:1998raa,Buras:1998raa,Buras:2003zz,Buras:1996dq}, 
\begin{multline}
\label{eq:Hamiltonian}
\mathcal{H}_{\rm eff.}^{\Delta S=1} =\\
- \frac{G_F V_{us}^* V_{ud}}{\sqrt{2}} \left[ C_-(\mu^2) Q_-(\mu^2) + C_7(\mu^2)Q_7\right]\;,
\end{multline}
where $C_{-}(\mu^2)$ and $C_{7}(\mu^2)$ are the Wilson coefficients (see Appendix \ref{sec:AppendixCC} for their expressions) associated to the four-quark operators $Q_{-}(\mu^2)$ and $Q_7$ respectively, given by
\begin{align}
Q_-&=4(\bar{s}_L\gamma^\nu u_L)(\bar{u}_L\gamma_\nu d_L) - 4(\bar{s}_L\gamma^\nu d_L)(\bar{u}_L\gamma_\nu u_L)\label{eq:Q-},\\
Q_7&=2 \alpha(\bar{s}_L\gamma^\nu d_L)(\bar{e}\gamma_\nu e)\;.\label{eq:Q7}
\end{align}
The SD amplitude then takes the form,
\begin{multline}
\label{eq:AmplitudeSD}
\mathcal{A}\left(K\rightarrow \pi \ell^+ \ell^-\right) = - \frac{G_F V_{us}^* V_{ud}}{\sqrt{2}} \\
\times\braket{\pi \ell^+ \ell^-}{C_-(\mu^2) Q_-(\mu^2) + C_7(\mu^2)Q_7}{K}\,.
\end{multline}
Both, the Wilson coefficients and the four-quark operators depend on the renormalization scale $\mu$ that separates the two regimes. Nevertheless, the physical amplitude cannot depend on $\mu$. $Q_{7}$ in eq.~\eqref{eq:AmplitudeSD} is a $\mu^2$-independent operator, so that in order the amplitude to be $\mu^2$ independent, the Wilson coefficient $C_{7}(\mu^2)$  has to cancel the $\mu^2$ dependence in $C_{-}(\mu^2)Q_{-}(\mu^2)$. Some of the consequences of this SD property will be considered in a model independent form in Ref. \cite{Knecht}.

\section{The Bardeen-Buras-G\'erard framework}
\label{sec:BBG}

In Ref.~\cite{Buras:1985yx,Bardeen:1986vp,Bardeen:1986uz,Bardeen:1986vz,Hambye:1998sma,Gerard:1990dx,Buras:2014maa},  the authors use an order $p^2$  chiral Lagrangian and a physical cut-off $M$ to regularize the contributions beyond tree level instead of the usual local counter-terms (\textit{e.g.} the $L_i$ and $N_i$ constants). Consequently, their results exhibit a quadratic dependence on the physical cut-off $M$ which according to them  is a crucial ingredient in the matching of the meson and quark pictures. They argue that one can obtain a parametrization of non-perturbative QCD effects  by matching a low-energy Lagrangian, valid up to the scale $M$, to the logarithmic behaviour of relevant Wilson coefficients at high-energy.  In this work we refer to this computational method as the {\it Bardeen-Buras-G\'erard framework} (BBG).
 
In this context, the function $W_+(z)$ becomes a function of $q^2$ and $M^2$, 
\begin{equation}
\label{eq:W+defBBG}
W_+(z) \longmapsto W_+(z,M^2)\;.
\end{equation}
Our goal is to predict the values of the $a_+$ and $b_+$ coefficients using BBG framework.

At the matching scale $M$, the description for low and high energy must coincide; this means that the LD quadratic divergence in $M$ has to be numerically equal to the SD logarithmic divergence. Therefore, at $\mu^2=M^2$ the SD Hamiltonian,
\begin{multline}
\label{eq:HamiltonianBBG}
\mathcal{H}_{\rm eff.}^{\Delta S=1} =- \frac{G_F V_{us}^* V_{ud}}{\sqrt{2}} \\
\times\left[ C_-(M^2) Q_-(M^2) + C_7(M^2)Q_7\right]\;,
\end{multline}
must coincide with its chiral representation at LD.

\subsection{Amplitude properties}

The BBG approach considers only a chiral $\mathcal{O}(p^2)$ effective Lagrangian below the scale $M$, so that, since the loop calculations are regularized by the cut-off $M$, higher order Lagrangians (\textit{i.e.} with $L_i$ and $N_i$ constants) do not appear at all. Following their prescriptions, one has then,
\begin{multline}
\label{eq:AmplitudeCQ}
\mathcal{A}\left(K\rightarrow \pi \ell^+ \ell^-\right) = - \frac{G_F V_{us}^* V_{ud}}{\sqrt{2}} \\
\times\braket{\pi \ell^+ \ell^-}{C_-(M^2) Q_-(M^2) + C_7(M^2)Q_7}{K}\,.
\end{multline}

The chiral loop calculation of the matrix element of $Q_-$ with the $\mathcal{O}(p^2)$ chiral Lagrangian does not provide any quadratic divergences (of course not in dimensional regularization) even in the cut-off regularization (see Appendix \ref{sec:AppendixAmplitudes}). The $\ln M^2$ appearing here at the chiral scale, is cancelled by local counter-terms in eq. \eqref{eq:w+} in usual $\chi$PT. Now, this role is played by $C_7(\mu^2=M^2)Q_7$.

\begin{figure}[h]
\begin{center}
\begin{tikzpicture}[scale=1]
   \begin{scope}[thick]
    \draw [->] (-0.5,0) -- (5,0) node [below right]  {$\mu^2$};
   \draw[gray,dotted] (0,3.2) -- (0,-0.2)  node [below] {$0$};    
    \draw[gray,dotted] (2.4,3.2) -- (2.4,-0.2)  node [below] {$M^2$};
   \end{scope}
   \draw [fill=blue, opacity=0.2] (0,0) rectangle (2.4,3);
   \draw [fill=green, opacity=0.2] (2.42,0) rectangle (4.8,3);
   \draw[->, >=latex, blue, line width=5pt] (0.1, 1) -- (2.3, 1);
   \draw[->, >=latex, green, line width=5pt] (4.8, 1) -- (2.5, 1);
\node at (1.2,2) [] {\blue Evolution Op.};
\node at (1.2,1.5) [] {\blue $\mathcal{E}(M^2)$};
\node at (3.6,2.5) [] {\red Usual Wilsonian};
\node at (3.7,2) [] {\red 4-quarks Op.};
\node at (3.7,1.5) [] {\red  RG Flow};
\end{tikzpicture}
\end{center}
\caption{The usual Wilsonian Renormalization flow is represented above the scale $M^2$. The extended Renormalization flow defined in eq. \eqref{eq:evolutionOp} is shown below $M^2$.}\label{fig:Sectors}
\end{figure}
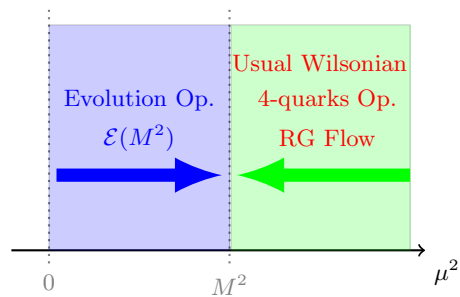

The matching between SD and LD should be around 1 GeV, then  $C_-(M^2) Q_-(M^2)$ and $C_7(\mu^2=M^2)Q_7$ have to evolve from the chiral scale to 1 GeV. But this evolution implies a mixing between the operators $Q_-$ and $Q_7$ according to RGE \cite{Gilman_Wise,Buras:1991jm,Buras:1992tc,Buras:1992zv,Buchalla:1995vs,Buras:1993dy,Buras:1998raa,Buras:1998raa,Buras:2003zz,Buras:1996dq}. In the BBG framework, this mixing is captured by the quadratic divergences \cite{Buras:1985yx,Bardeen:1986vp,Bardeen:1986uz,Bardeen:1986vz,Hambye:1998sma,Gerard:1990dx,Buras:2014maa} which in our case can come only from the $K\rightarrow 3\pi$ vertex (chiraly related to $K\rightarrow 2\pi$ studied by BBG see below). In other words, the authors of Ref.\cite{Buras:1985yx,Bardeen:1986vp,Bardeen:1986uz,Bardeen:1986vz,Hambye:1998sma,Gerard:1990dx,Buras:2014maa} have extended the usual renormalization flow of the SD sector (from $M^2_W$ to $M^2$) to a flow in the LD sector from $M^2$ to $0$ (as depicted in Fig. \ref{fig:Sectors}) through the relation,
\begin{equation}
Q_-(M^2) = \mathcal{E}(M^2) Q_-(0)\;, 
\end{equation}
where $ \mathcal{E}(M^2)$ is the \textit{evolution operator} given by \cite{Buras:2014maa},
\begin{equation}
\label{eq:evolutionOp}
\mathcal{E}(M^2) \doteq 1 + \frac{3}{16\pi^2} \left[\frac{M^2}{f_\pi^2} + \frac{M_K^2}{4f_\pi^2}\ln\left(1+\frac{M^2}{\tilde{m}^2}\right)\right]\;,
\end{equation}
with $\tilde{m}\approx 0.3$ GeV. $\mathcal{E}(M^2)$ comes from the $K\rightarrow \pi\pi$ analysis in Ref. \cite{Buras:2014maa}, and soft-pion theorem tells us that it can be applied to $K\rightarrow 3\pi$ vertex~\cite{deRafael:1995zv,Pich:1995bw,Ecker:1995zu,D'Ambrosio:1996nm,Komatsubara:2012pn,Cirigliano:2011ny,Littenberg:2002um,Colangelo:2000zw}. The amplitude is then given by
\begin{multline}
C_-(M^2) \braket{3\pi }{Q_-(M^2)}{K} =\\
C_-(M^2)\mathcal{E}(M^2) \braket{3\pi}{Q_-(0)}{K}\;.
\end{multline}

The authors of Ref. \cite{Buras:2014maa} find that the range of numerical values for $M$ that leaves the amplitude invariant is, 
\begin{equation}
0.6 \text{ GeV} \leqslant M < 1 \text{ GeV}\;,
\end{equation}
with a preferred value at $0.7$ GeV (without vector contribution).

\section{Determination of $\boldsymbol{a_+}$ and $\boldsymbol{b_+}$ (no vectors)}

Eq.~\eqref{eq:Wexpansion} determines uniquely the coefficients $a_+$ and $b_+$ as we will see here. Writing
\begin{equation}
W_+(z,M^2)\underset{z \rightarrow 0}{\thicksim} M_K^2 G_Fa_+(M^2) + M_K^2 G_Fb_+(M^2) z \;,
\end{equation}
we identify (the wave function renormalization factors $Z_\pi$ and $Z_K$ are given in App. \ref{sec:AppendixAmplitudes}),
\begin{align}
&a_+(M^2) =\nonumber\\
&-\frac{V_{us}^* V_{ud}}{\sqrt{2}}\sqrt{Z_\pi Z_K} \bigg\{-4\pi C_7(M^2) \nonumber\\
&\hspace*{0.5cm}+ C_-(M^2) \left[- \frac{5}{9} + \frac{1}{3} \ln\frac{M^2}{M_\pi M_K} \right] \mathcal{E}(M^2) \bigg\} \label{eq:a+}\;.
\end{align}

\begin{figure}[h]
\begin{center}
\begin{tikzpicture}[scale=0.8]
\begin{axis}[xlabel={$M$  [GeV]},ylabel={$a_+$},ymin=-0.6,ymax=0,xmin=0.6,xmax=1,legend pos=north east]
\addplot[dotted,green!60!black,line width=1.5pt] table {BBGCminus.dat};
\addplot[dashed,orange,line width=1.5pt] table {BBGC7.dat};
\addplot[blue,line width=1.5pt] table {aplus.dat};
\addplot[blue,dashed,line width=1pt] coordinates { (0.7,-0.8) (0.7,0.0)};
\legend{$C_-$,$C_7$,$a_+$}
\end{axis}
\end{tikzpicture}
\caption{ In blue, the variation of $a_+$ as a function of $M$ in GeV.  The dotted green curve represents the contribution proportional to $C_-(M^2)$ and the dashed orange curve the one proportional to $C_7(M^2)$. The vertical dashed line stands for the matching scale.} \label{fig:aplus}
\end{center}
\end{figure}
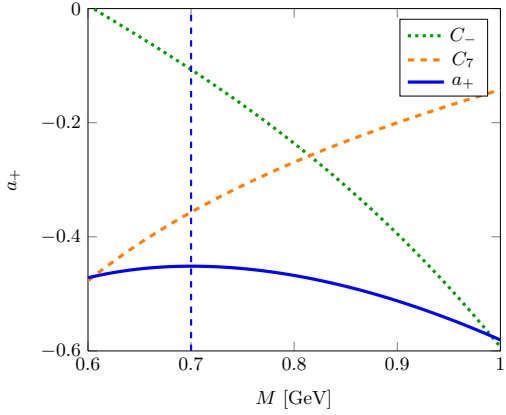

Compared to the analysis of $K\rightarrow 2\pi$ in Ref.\cite{Buras:1985yx,Bardeen:1986vp,Bardeen:1986uz,Bardeen:1986vz,Hambye:1998sma,Gerard:1990dx,Buras:2014maa}, we have additionally a further  cancellation of the log in  $\left[- \frac{5}{9} + \frac{1}{3} \ln\frac{M^2}{M_\pi M_K} \right]$ and the log in $C_7(M^2)$. This fixes $M$ and then $a_+$ as shown in Fig. \ref{fig:aplus}. We have also, 
\begin{align}
&b_+(M^2) = -\frac{V_{us}^* V_{ud}}{\sqrt{2}} \frac{\sqrt{Z_\pi Z_K}}{60r_\pi^2} C_-(M^2) \mathcal{E}(M^2) \nonumber \\
&\hspace*{2cm}-\frac{1}{M_K^2} \frac{3r_\pi^2(\alpha_+ - \beta_+) - \beta_+}{180G_F r_\pi^6} \;.
\end{align}

In order to find the value of $M$ where there is a compensation between the LD quadratic dependence (including both terms in eq.\eqref{eq:a+}, constant and the novel logarithmic one)  and the SD logarithm, we look for the solution of $\partial_{M^2} a_+ =0$. We find that this equation is satisfied when $M=0.7$ GeV and numerically one gets  
\begin{align}
a_+\left((0.7 \text{ GeV})^2 \right) &= -0.5, \\ 
b_+\left((0.7 \text{ GeV})^2 \right) &= -0.12\;.
\end{align}

Comparing with the experimental values eq. \eqref{eq:exp}, we find a good agreement for $a_+$, but not for $b_+$.  Fig.~\ref{fig:aplus} shows $a_+$ as a function of $M$, together with the contributions coming from  $C_-$ and  $C_7$, separately.  These are the expected behaviours from LD physics. The dashed vertical line corresponds to the scale where $\partial_{M^2} a_+ =0$.

In the following section we study the inclusion of vectors in the BBG approach.  


\section{Vector contributions in the BBG framework}

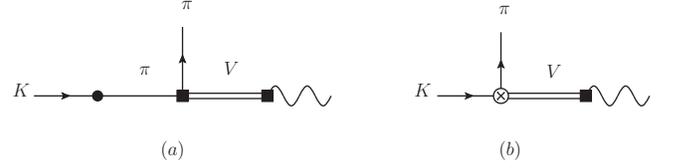
\begin{figure}[h]
\begin{center}
\scalebox{0.5}{ \begin{picture}(478,130) (51,-9)
    \SetWidth{0.5}
    \SetColor{Black}
    \Text(160,-14)[lb]{\Large{\Black{$(a)$}}}
    \SetWidth{1.0}
    \Line[arrow,arrowpos=0.5,arrowlength=5,arrowwidth=2,arrowinset=0.2](64,34)(112,34)
    \Line[arrow,arrowpos=0.5,arrowlength=5,arrowwidth=2,arrowinset=0.2](176,38)(176,86)
    \Line(112,34)(176,34)
    \Photon(240,34)(288,34){7.5}{2}
    \SetWidth{0.0}
    \Vertex(112,34){4.243}
    \Text(48,34)[lb]{\Large{\Black{$K$}}}
    \Text(176,100)[lb]{\Large{\Black{$\pi$}}}
    \Text(144,50)[lb]{\Large{\Black{$\pi$}}}
    \CBox(180.243,29.757)(171.757,38.243){Black}{Black}
    \Text(416,-14)[lb]{\Large{\Black{$(b)$}}}
    \SetWidth{1.0}
    \Line[arrow,arrowpos=0.5,arrowlength=5,arrowwidth=2,arrowinset=0.2](368,34)(416,34)
    \Line[arrow,arrowpos=0.5,arrowlength=5,arrowwidth=2,arrowinset=0.2](416,34)(416,82)
    \Photon(480,34)(528,34){7.5}{2}
    \Text(352,34)[lb]{\Large{\Black{$K$}}}
    \Text(416,96)[lb]{\Large{\Black{$\pi$}}}
    \Text(452,48)[lb]{\Large{\Black{$V$}}}
    \Line[double,sep=4](416,34)(480,34)
    \COval(416,34)(5.657,5.657)(45.0){Black}{White}\Line(413.172,31.172)(418.828,36.828)\Line(413.172,36.828)(418.828,31.172)
    \SetWidth{0.0}
    \CBox(484.243,29.757)(475.757,38.243){Black}{Black}
    \SetWidth{1.0}
    \Line[double,sep=4](176,34)(240,34)
    \SetWidth{0.0}
    \CBox(244.243,29.757)(235.757,38.243){Black}{Black}
    \Text(208,50)[lb]{\Large{\Black{$V$}}}
  \end{picture}
}
\end{center}
\caption{Type $(a)$ diagram represents the VMD contribution to $L_9$ ( the $\bullet$ vertex is $\mathcal{O}(p^2)$ $\Delta S =1$ vertex). Analogously,  type $(b)$ diagram represents the VMD contribution to the $N_{14}-N_{15}$ one ( the $\otimes$ vertex is $\Delta S =1$ vertex coming from $Q_-$). \label{fig:vectincl}}
\end{figure}

Vector contributions increase the range of validity of $M^2$ and smooth over the transitions between short and long distance continuation \cite{Buras:1985yx,Bardeen:1986vp,Bardeen:1986uz,Bardeen:1986vz,Hambye:1998sma,Gerard:1990dx,Buras:2014maa}. We have to consider two  counter-term structures from eq. \eqref{eq:w+} which are shown in Fig. \ref{fig:vectincl}, the structure coming from $L_9$ (diagrams $(a)$) and  the $N_{14}-N_{15}$ contributions (diagrams $(b)$ ). 

The diagrams $(a)$ drive to the inclusion of vectors with mass $M_V$ in the evolution operator of $Q_-$ in eq.~\eqref{eq:evolutionOp} as explained in \cite{Buras:2014maa}, 
\begin{equation}
\label{eq:evolopvect}
\mathcal{E}(M^2) \longmapsto \mathcal{E}(M^2,M_V^2)= \mathcal{E}(M^2) + \Delta(M^2,M_V^2)\;, 
\end{equation}
where
\begin{multline}
\Delta(M^2,M_V^2) = \frac{3}{16\pi^2} \bigg[-\frac{9}{16}\frac{M^2}{f_\pi^2} +\frac{3}{8}\frac{M^2}{f_\pi^2}\frac{M_V^2}{{M^2}+{M_V^2}}\\
+ \frac{3}{16}\frac{M_V^2}{f_\pi^2}\ln\left(1+\frac{M^2}{M_V^2}\right)\bigg],\;
\end{multline}
and change the electromagnetic form factor 
\begin{equation}
1 \longmapsto 1+ z\frac{M_K^2}{M_V^2}\;.
\end{equation}

The diagrams $(b)$ in Fig. \ref{fig:vectincl} corresponding to the $N_{14}-N_{15}$ local counter-terms imply a modification of the mixing between $Q_-$ and $Q_7$ in the RGE by adding an extra contribution $C_-(M^2) \eta_V(M^2,z)$. This contribution is not present in  $K\rightarrow \pi\pi$ processes and so does not affect the results in Ref. \cite{Buras:1985yx,Bardeen:1986vp,Bardeen:1986uz,Bardeen:1986vz,Hambye:1998sma,Gerard:1990dx,Buras:2014maa}. The complete calculation with vectors can be done using the Hidden Local Symmetry framework \cite{Bando:1984pw,Bando:1984ej,Bando:1985rf,Ecker:1989yg,Harada:2003jx}. We have to be careful, the counting in Large-N$_c$ must be respected by including all terms up to $\sfrac{1}{N_c}$ corrections with the same argument in Sec. \ref{sec:BBG}. One can evaluate this contribution as 
\begin{equation}
\eta_V(M^2,z) = 4 \pi \left[ \frac{f_\pi^2}{M_V^2} - z \frac{2}{3} \frac{M_K^2}{M_V^2} \ln \frac{M^2}{M_K^2}\right]\;.
\end{equation}

One gets therefore, 
\begin{align}
&a_+(M^2,M_V^2) =\nonumber\\
&-\frac{V_{us}^* V_{ud}}{\sqrt{2}}\sqrt{Z_\pi Z_K} \bigg\{-4\pi C_7(M^2) \nonumber\\
&\hspace*{0.5cm}+ C_-(M^2) \left[- \frac{5}{9} + \frac{1}{3} \ln\frac{M^2}{M_\pi M_K} \right] \mathcal{E}(M^2,M_V^2)\nonumber \\
&\hspace*{0.5cm}+ C_-(M^2) \; 4\pi \; \frac{f_\pi^2}{M_V^2} \bigg\} \label{eq:aplusvect2},
\end{align}
and
\begin{align}
&b_+(M^2,M_V^2) = \nonumber\\
&\frac{M_K^2}{M_V^2} a_+(M^2,M_V^2)-\frac{1}{M_K^2} \frac{3r_\pi^2(\alpha_+ - \beta_+) - \beta_+}{180 G_F r_\pi^6}\nonumber\\
&\hspace*{0.2cm}-\frac{V_{us}^* V_{ud}}{\sqrt{2}}\sqrt{Z_\pi Z_K} C_-(M^2) \bigg[\frac{1}{60r_\pi^2} \mathcal{E}(M^2,M_V^2)\nonumber\\
&\hspace*{4.8cm}- \frac{8\pi}{3} \frac{M_K^2}{M_V^2}\ln \frac{M^2}{M_K^2}\bigg] \;.
\end{align}

In the same manner as before, we evaluate the scale $M$ by requiring $\partial_{M^2} a_+ =0$ in eq. (\ref{eq:aplusvect2}), and obtain that for  $M=0.7$ GeV  
\begin{align}
a_+\left((0.7 \text{ GeV})^2,(0.775 \text{ GeV})^2 \right) &= -0.54, \\ 
b_+\left((0.7 \text{ GeV})^2,(0.775  \text{ GeV})^2 \right)&= -0.72\;.
\end{align}

The interplay between strong amplitudes ($L_9$) with external weak transitions (diagrams $(b)$ in Fig.\ref{fig:vectincl}) have been already noticed by the authors of Ref.\cite{Ecker:1990in}  for the VMD $\mathcal{O}(p^6)$ contribution to $K_L\rightarrow \pi^0 \gamma\gamma$.

\begin{figure}[h]
\begin{center}
\begin{tikzpicture}[scale=0.8]
\begin{axis}[xlabel={$M$ [GeV]},ylabel={$a_+$},ymin=-0.7,ymax=-0.4,xmin=0.6,xmax=1,legend cell align=left,legend pos= north east]
\addplot[blue,line width=1.5pt] table {aplusvecteta.dat};
\addplot[blue,line width=1.5pt,dotted] table {aplusvect.dat};
\addplot[blue,line width=1.5pt,dashed] table {aplus.dat};
\addplot[blue,dashed,line width=1pt] coordinates { (0.7,-0.8) (0.7,4)};
\legend{BBG(vect) $(a)+(b)$, BBG no vect, BBG(vect)(a)}
\end{axis}
\end{tikzpicture}
\caption{ $a_+$ as a function of $M$ in the three different frameworks: `BBG no vect.' where vectors are not included, `BBG(vect)$(a)$'~ represents the contribution coming only from diagrams $(a)$ in Fig. \ref{fig:vectincl} and  `BBG(vect) $(a)+(b)$'~ is the case where both $(a)$ and $(b)$ diagrams were included.   The vertical line indicates the value $M=0.7$ GeV.\label{fig:3aplus} } 
\end{center}
\end{figure}
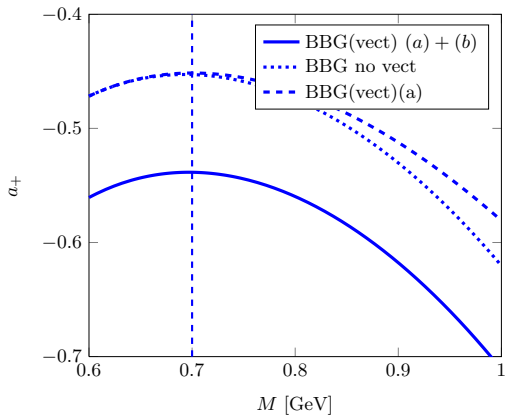

We show in Fig.~\ref{fig:3aplus}, $a_+$ as a function of $M$ in the three different scenarios: `BBG no vect.'~is the framework where no vectors are included at all and `BBG(vect)$(a)$'~is the one where only the diagrams $(a)$ in Fig. \ref{fig:vectincl} are considered. We refer to `BBG(vect) $(a)+(b)$'~as the last case where all kinds of diagrams in Fig. \ref{fig:vectincl} have been included.

Following Buchalla \textit{et al.} in Ref.~\cite{Buchalla:2003sj}, we investigate our predictions of what the authors call $a_+^\text{VMD}$ and $a_+^\text{nVMD}$. Under the general hypothesis that the $b_+$ term in eq.~\eqref{eq:W+decomp} is generated by the expansion  of a vector-meson propagator, $W_{+}(z)$ can be written as,
\begin{align}
W_+(z) = G_F M_K^2 \bigg[(a_{+}^\text{VMD}+a_{+}^\text{nVMD})+a_{+}^\text{VMD}\frac{M_K^2}{M_V^2}z\bigg]\;,
\end{align}
where $a_{+}^\text{nVMD}$ denotes $z$-independent non-VDM contributions. The introduction of the $\eta_V$ contribution is necessary to recover this separation between $a_+^\text{VMD}$ and $a_+^\text{nVMD}$. Indeed, we find,
\begin{align}
a_+^\text{VMD} &= \frac{M_V^2}{M_K^2}\left(b_+\big\vert_{\text{BBG(vect)}(a)+(b)} - b_+\big\vert_{\text{BBG}} \right) \\
&=  \frac{M_V^2}{M_K^2}\left[-0.72 - (-0.12) \right] = -1.5 \;,
\end{align}
which is in good agreement with $a_+^\text{VMD} =\frac{M_K^2}{M_V^2}b_{+}^{\text{exp}}=-1.6\pm0.1$ \cite{Buchalla:2003sj}. 

\section{Analysis of $K_S \rightarrow \pi^0 \ell^+ \ell^-$}
\label{sec:KS}

The analysis of $K_S \rightarrow \pi^0 \bar\ell \ell$ can be directly deduced from the previous one, $K^+ \rightarrow \pi^+ \bar\ell \ell$. Indeed, 
\begin{multline}
\braket{\pi^0 \gamma^*(q)}{Q_-(0)}{K_S} \\
= -  \braket{\pi^0 \gamma^*(q)}{Q_-(0)}{K^+}\bigg\vert_{M_\pi=M_K}\;.
\end{multline}

And, in this case, the local counter-term structures at $\mathcal{O}(p^4)$ are~\cite{D'Ambrosio:1998yj}
\begin{equation}
a_s^{(4)} = \frac{G_8}{G_F}\left( \frac{1}{3}-w_s\right)\;,
\end{equation} 
where $w_s$ is \cite{Ecker:1992de,Gasser:1984pr,Kpee_LD}
\begin{equation}
w_S  =  \frac{32 \pi^2}{3} \left( 2 N_{14}^r + N_{15}^r \right) + \frac{1}{3} \ln \frac{\mu^2}{M_K^2}.
\end{equation}

Given the decay $K_S \rightarrow 3\pi$ ($\Delta I=\sfrac{1}{2}$ transitions) is not allowed ($\Delta I=\sfrac{3}{2}$ transitions are permitted), only kaons are present in the loop (see Appendix \ref{sec:AppendixAmplitudes}). Using the same identification as in eq.~\eqref{eq:Wexpansion} and following the same procedure as in the case of the decay $K^+ \rightarrow \pi^+ \bar\ell \ell$, we find that $a_S=1.2$ $(a_S^\text{exp}=|1.08|^{+0.26}_{-0.21}$ \cite{NA48/2} ) for the same scale $M=0.7$ GeV established from eq. (\ref{eq:aplusvect2}). This value is in agreement with the fitted $w_S$ value obtained in Ref. \cite{Friot:2004yr}.


\section{Conclusion}

We have evaluated the $K^+\rightarrow \pi^+ \ell^+ \ell^- $ form factor parameters  $a_+$ and $b_+$ in the BBG framework. Regarding $a_+$ the theoretical dependence/uncertainty in this framework on the matching scale seems small, see Fig.3: comparison with phenomenology seems very successful, see eq. (24).
Consistency with the full chiral structure of the weak counter-terms has required a more general discussion on vector contributions (see section VI and Fig.4) that leads to an extension of the $Q_-$ evolution studied by the authors of Ref. \cite{Buras:1985yx,Bardeen:1986vp,Bardeen:1986uz,Bardeen:1986vz,Hambye:1998sma,Gerard:1990dx,Buras:2014maa} in the context of $K\rightarrow 2\pi$. This extension met nicely with the experimental values \cite{NA48/2}. We have applied our method to $K_S \rightarrow \pi^0 e^+ e^- $ in section VII and found a good agreement with experimental results too.


\begin{acknowledgements}

The authors are grateful to the Mainz Institute for Theoretical Physics (MITP) for the hospitality and partial support during the completion of this work. We thank A. Buras,  E. de Rafael, J.-M. G\'{e}rard, T. Hambye, C. Sachrajda for discussions and M. Knecht for valuable comments and further collaboration. D.G. were supported in part by MIUR under project 2010YJ2NYW and by INFN research initiative PhenoLNF.

\end{acknowledgements}

\appendix

\section{The evaluation of scalar integrals}
\label{sec:AppendixSacalrIntegral}

The loop integral with a cut-off $M^2$ has the form,
\begin{equation}
I(\alpha,R^2,M^2) \doteq -i \int^{M^2}\! \frac{\d^4 \ell}{(2\pi)^4} \frac{1}{\left(\ell^2-R^2\right)^\alpha}\;,
\end{equation}
and gives
\begin{equation}
\label{eq:scalarIntegral}
I(\alpha,R^2,M^2) =\frac{M^4 R^{-2\alpha}}{6(4\pi)^2}  \hyper{2}{1}{\alpha,2}{3}{-\frac{M^2}{R^2}}\;, 
\end{equation}
$_2F_1$ is the Gauss' hypergeometric function. In the one loop case for example, the integral is given by 
\begin{align}
\mathrm{A}_0\left(m^2\right)&=I(1,m^2,M^2)\nonumber\\
&= \frac{1}{3(4\pi)^2}\left[ M^2 - m^2 \ln\left(1+\frac{M^2}{m^2}\right)\right]\;.
\end{align}
All the scalar integrals can be evaluated using eq.~\eqref{eq:scalarIntegral}.

\section{Amplitudes formulae}
\label{sec:AppendixAmplitudes}

\subsection{$\boldsymbol{K^+\rightarrow \pi^+\gamma^*}$}

The form factor defined in eq.~\eqref{eq:W+defBBG} is obtained from 
\begin{multline}
W_+(z,M^2) = \frac{M_K^2 G_F V_{us}^* V_{ud}}{\sqrt{2}}\sqrt{Z_\pi Z_K}\times\\
\left[ C_-(M^2) \braket{\pi^+ \gamma^*(q) }{Q_-(M^2)}{K^+} + 4\pi C_7(M^2) \right]\;,
\end{multline}
where,
\begin{multline}
\braket{\pi^+ \gamma^*(q) }{Q_-(M^2)}{K^+} \\
=\mathcal{E}(M^2)\braket{\pi^+ \gamma^*(q)}{Q_-(0)}{K^+}\;,
\end{multline}
and
\begin{multline}
\sqrt{Z_\pi Z_K} = 1 + \frac{1}{16\pi^2}\bigg[\frac{M^2}{f_\pi^2}-\frac{5}{12}\frac{M_K^2}{f_\pi^2}\ln\left(1+\frac{M^2}{M_K^2}\right) \\
-\frac{1}{8}\frac{M_\eta^2}{f_\pi^2}\ln\left(1+\frac{M^2}{M_\eta^2}\right) \bigg]\;.
\end{multline}
From a pure $\chi$PT loop calculation using the cut-off prescription in eq.~\eqref{eq:scalarIntegral}, one has
\begin{multline}
\label{eq:pi+Q-0K+}
\braket{\pi^+ \gamma^*(q)}{Q_-(0)}{K^+} \\
= \chi\left(\frac{z}{r_\pi^2}\right) +\chi\left(z\right) - \frac{5}{9} + \frac{1}{3} \ln\frac{M^2}{M_\pi M_K}\;.
\end{multline}

The $\chi$ function is the one defined in Ref.~\cite{D'Ambrosio:1998yj} and it is related to the $\Phi$ in Ref.~\cite{Kpee_LD} as $\chi(z)= \Phi(z) + \sfrac{1}{6}$. Numerically the kaon loop contribution, the $\chi(z)$ term in eq.~\eqref{eq:pi+Q-0K+}, is negligible. The extra constant term and $\ln(M^2)$ in~\eqref{eq:pi+Q-0K+} come from the cut-off regularization. It is from this formula that one can extract the expressions for $a_+(M^2)$ and $b_+(M^2)$.

\subsection{$\boldsymbol{K_S\rightarrow \pi^0\gamma^*}$}

For this decay, the form factor $W_S(z,M^2)$ is,
\begin{multline}
W_S(z,M^2) = \frac{M_K^2 G_F V_{us}^* V_{ud}}{\sqrt{2}}\sqrt{Z_\pi Z_K}\times\\
\left[ C_-(M^2) \braket{\pi^0 \gamma^*(q) }{Q_-(M^2)}{K_S} - 4\pi C_7(M^2) \right]\;.
\end{multline}
The evolution operator in eq.~\eqref{eq:evolutionOp} is exactly the same as in the $K_S$ case, so, 
\begin{multline}
\braket{\pi^0 \gamma^*(q) }{Q_-(M^2)}{K_S} \\
=\mathcal{E}(M^2)\braket{\pi^0 \gamma^*(q)}{Q_-(0)}{K_S}\;,
\end{multline}
where,
\begin{multline}
\label{eq:pi0Q-0KS}
\braket{\pi^0 \gamma^*(q)}{Q_-(0)}{K_S} \\
= 2\chi\left(z\right) - \frac{5}{9} + \frac{1}{3} \ln\frac{M^2}{M_K^2}\;.
\end{multline}

\section{Expressions for $\boldsymbol{C_-(\mu^2)}$ and $\boldsymbol{C_7(\mu^2)}$}
\label{sec:AppendixCC}
The expressions for $C_-(\mu^2)$ and $C_7(\mu^2)$ are~\cite{Gilman_Wise},
\begin{equation}
C_-(\mu^2)= \frac{1}{2}\left[\frac{\alpha_s(\mu^2,4)}{\alpha_s(M_c^2,3)}\right]^{\frac{12}{27}}\left[\frac{\alpha_s(M_c^2,4)}{\alpha_s(M_W^2,4)}\right]^{\frac{12}{25}},
\end{equation}
and
\begin{multline}
C_7(\mu^2)=\frac{16}{99\alpha_s(M_c^2,3)} \\
\times\bigg\{ \left[\frac{\alpha_s(M_c^2,4)}{\alpha_s(M_W^2,4)}\right]^{-\frac{6}{25}}\bigg[1 -\left(\frac{\alpha_s(\mu^2,3)}{\alpha_s(M_c^2,3)} \right)^{-\frac{33}{27}}\bigg]\\
-\frac{11}{10}\left[\frac{\alpha_s(M_c^2,4)}{\alpha_s(M_W^2,4)}\right]^{\frac{12}{25}}\bigg[1 -\left(\frac{\alpha_s(\mu^2,3)}{\alpha_s(M_c^2,3)} \right)^{-\frac{15}{27}}\bigg]\bigg\},
\end{multline}
where
\begin{equation}
\alpha_s(\mu^2,n) = \frac{12 \pi}{33-2n} \frac{1}{\ln\left(\frac{\mu^2}{(0.3 \text{GeV})^2}\right)}\;.
\end{equation}


\end{document}